\documentclass[12pt]{article}

\usepackage{graphicx}
\usepackage{epsfig}
\usepackage{amsmath}
\usepackage{amssymb}
\usepackage{amscd}
\usepackage{amsfonts}
\usepackage{appendix}
\usepackage{bbold}
\usepackage{cite}
\usepackage[footnotesize]{caption2}
\usepackage{graphicx}
\usepackage[center,footnotesize,hang]{subfigure}
\usepackage{url}
\usepackage{dcolumn}
\usepackage{bbm}
\usepackage{lscape}

\usepackage{a4}
\usepackage{a4wide}
\usepackage{wasysym}

\newcommand{\Rep}[1]{\underline{\mbox{\textbf{#1}}}}

\begin{document}
\begin{center}
{\Large\sffamily\bfseries
\mathversion{bold} Non-Abelian Discrete Groups from the Breaking of Continuous Flavor Symmetries
\mathversion{normal}}
\\[13mm]
{\large
A.~Adulpravitchai~\footnote{E-mail: \texttt{adisorn.adulpravitchai@mpi-hd.mpg.de}}, 
A.~Blum~\footnote{E-mail: \texttt{alexander.blum@mpi-hd.mpg.de}} and
M.~Lindner~\footnote{E-mail: \texttt{manfred.lindner@mpi-hd.mpg.de}}} 
\\[5mm]
{\small \textit{ 
Max-Planck-Institut f\"{u}r Kernphysik\\ 
Postfach 10 39 80, 69029 Heidelberg, Germany
}}
\vspace*{1.0cm}
\end{center}
\normalsize
\begin{abstract}
\noindent We discuss the possibility of obtaining a non-abelian discrete flavor symmetry from an underlying continuous, possibly gauged, flavor symmetry $SU(2)$ or $SU(3)$ through spontaneous symmetry breaking. We consider all possible cases, where the continuous symmetry is broken by small representations. ``Small'' representations are these which couple at leading order to the Standard Model fermions transforming as two- or three-dimensional representations of the flavor group. We find that, given this limited representation content, the only non-abelian discrete group which can arise as a residual symmetry is the quaternion group $D_2'$.
\end{abstract}

\section{Introduction}

Ever since the discovery of large neutrino mixing, non-abelian discrete flavor symmetries have been a popular and quite successful approach towards describing the mixing patterns of the Standard Model of Particle Physics (SM). Based on the success of symmetries, it is well-motivated to invent a new global flavor symmetry as explanation for the existence of generations. The breaking of such symmetry could then justify phenomenological successful non-abelian discrete symmetries as unbroken subgroups. We want to take a closer look at the possible origin of discrete symmetries governing the structure of the SM Yukawa couplings. A natural scenario is to assume that the discrete flavor symmetry is connected to some underlying space-time or internal gauge symmetries. Connecting a flavor symmetry to the symmetries of space-time necessitates an extension of space-time itself. Thus flavor symmetries have been connected with discrete symmetries arising in compactified extra dimensions, with \cite{Kobayashi:2006wq} or without \cite{Altarelli:2006kg,Ourorbifold} string theory.

In this paper, we will consider the other possibility, i.e.\ that a discrete flavor symmetry is a conserved residual subgroup of a spontaneously broken gauged flavor symmetry. The idea of embedding a discrete flavor symmetry in a larger continuous group has been discussed in the literature, for example in \cite{subgroups,Hagedorn:2006ug}. However no complete models exist, in the sense that the discrete flavor symmetry is only motivated by a possible underlying continuous symmetry, but the Lagrangian used for phenomenological considerations is only invariant under the discrete group, i.e.\ the continuous group is explicitly broken such as for example in \cite{Tanimoto:1999pj}. This leads to restrictions on representations and to correlations between Yukawa couplings, not only through group theoretical compatibility, but also through demanding anomaly freedom for an underlying gauge symmetry \cite{Araki:2008ek,Luhn:2008sa}. This does not however solve the problem of the underlying symmetry breaking dynamics. (For the discussion of the underlying symmetry breaking from continuous symmetries to their continuous subgroups see \cite{Li:1973mq}.) 

To obtain a complete model, one needs to determine the scalar representations that break the gauge symmetry as well as their Vacuum Expectation Value (VEV) structure. In general the necessary representations are well known, and are in fact mentioned as a motivation in some discrete flavor symmetry models. The necessary VEV structures have also been partially discussed \cite{subgroups}, however often not in a flavor context \cite{Etesi:1997jv,Koca:1997td,Koca:2003jy}. Here the reasons for the absence of complete symmetry breaking models become clear: In general large and unwieldy representations are needed to break down to a phenomenologically interesting discrete subgroup. It is the aim of this paper to show that in fact the best known and simplest representations lead only to the conservation of a very limited amount of non-abelian discrete subgroups. In fact, we will show that the only non-abelian discrete subgroup with the small representations discussed is $D_2'$,\footnote{The dihedral group $D_2$ being an abelian group has four one-dimensional irreducible representations. The order of $D_2$ is 4. The double valued counterpart of the dihedral group $D_2$ is $D_2'$. $D_2'$ has four one- and one two-dimensional irreducible representations. The order of $D_2'$ is 8. $D_2'$ is the simplest non-abelian double valued dihedral group and is also called the quaternion group. For more information see \cite{Blum:2007jz}.} which has been used as a flavor symmetry \cite{Frigerio:2004jg}, but does not have a rich enough structure to predict by itself very specific mixing patterns, such as tri-bimaximal mixing for neutrinos \cite{Blum:2007jz}.

We want to break a hypothetical continuous flavor symmetry (gauged or not is irrelevant for this discussion) at a high energy scale. This flavor symmetry should commute with the SM and transform the three generations of fermions into each other. If we limit ourselves to three generations, we only need to consider the groups $SU(2)$ and $SU(3)$ as all other semi-simple Lie groups do not have two- or three-dimensional representations. We do not need to discuss an $SO(3)$ separately, since the $SO(3)$ gauge theory can simply be considered as an $SU(2)$ theory with a limited representation content. For $SU(2)$ the fermions will transform as $\Rep2 + \Rep1$ or $\Rep3$. The relevant Kronecker products are thus

\begin{eqnarray}
\Rep2 \times \Rep2 &=& \Rep1 + \Rep3, \nonumber\\
\Rep2 \times \Rep3 &=& \Rep2 + \Rep4, \\
\Rep3 \times \Rep3 &=& \Rep1 + \Rep3 + \Rep5. \nonumber
\end{eqnarray}

\noindent For a flavor symmetry $SU(3)$ the possible representations for three fermion generations are $\Rep3$ and $\Rep{$\overline{\bf 3}$}$, with Kronecker products %
\begin{eqnarray}
\Rep3 \times \Rep3 &=& \Rep{$\overline{\bf 3}$} + \Rep6, \nonumber\\
\Rep{$\overline{\bf 3}$} \times \Rep{$\overline{\bf 3}$} &=& \Rep3 + \Rep{$\overline{\bf 6}$},\\
\Rep3 \times  \Rep{$\overline{\bf 3}$} &=& \Rep1 + \Rep8. \nonumber
\end{eqnarray}

\noindent We will discuss the breaking of these flavor symmetries by small representations, where smallness means dimension not larger than five for $SU(2)$ and eight for $SU(3)$. This choice of representations is motivated by the fact that only these representations can couple to fermions at leading order, as can be read off from the Kronecker products above. The VEVs of these representations can easily be discussed using linear algebra. We discuss how the continuous symmetry is broken by VEVs of scalars transforming under these representations and show that no non-abelian discrete symmetries, apart from $D_2'$, can be conserved with these representations alone, and thus one generically needs larger representations and group theory beyond simple linear algebra to model such a scenario. In fact, $D_2'$ itself only arises as the double-valued group of the abelian $D_2$ if we break $SU(2)$ with the unfaithful five-dimensional representation, which is also a representation of $SO(3)$.

To determine whether a certain VEV structure conserves a subgroup of the flavor symmetry $SU(2)$ or $SU(3)$, we test which elements of the flavor symmetry leave the VEV invariant. We will assume the minimal scalar content for any representation, i.e.\ real scalars for real representations, complex scalars for pseudo-real and complex representations. We then check for each representation, which subgroups the VEV of a scalar field transforming under this representation can conserve. We also consider combinations of VEVs, but only where such a combination could lead to a non-abelian subgroup. We begin by discussing those representations which can be written with one index, i.e.\ which can be written as vectors in our linear algebra treatment. We continue with those representations with two indices, i.e.\ matrix representations. First we take the most familiar (from the SM gluons), the adjoint representation of $SU(3)$, then we continue with the very similar $\Rep5$ of $SO(3)$, $\Rep6$ of $SU(3)$ and $\Rep4$ of $SU(2)$ at the end of the paper. 

\section{Breaking Continuous symmetries}

\mathversion{bold}
\subsection{Breaking $SU(2)$ with a doublet}
\mathversion{normal}

In the two-dimensional representation of SU(2) the group elements are mapped onto the 2$\times$2 unitary matrices with unit determinant. Thereby each group element has two eigenvalues $\lambda_1$ and $\lambda_2$. They must obey the constraint that $\lambda_1 \lambda_2 = 1$, as the product of the eigenvalues is just the determinant. Hence if one of the eigenvalues is 1, then so is the other one. The only 2$\times$2 matrix with two eigenvalues of 1 is obviously the unit matrix. Hence, the identity element is the only element of the group that can leave a doublet VEV invariant. We conclude from this that the VEV of a scalar transforming as a doublet of $SU(2)$ always breaks the entire group. This of course does not change if we add further scalars of any sort.

\mathversion{bold}
\subsection{Breaking $SU(2)$ with a triplet}
\mathversion{normal}

The triplet is the fundamental representation of $SO(3)$ and an unfaithful representation of $SU(2)$. The group elements are mapped onto the $3\times3$ orthogonal matrices with unit determinant. These can be thought of as rotations in three-dimensional Euclidean space. If such a rotation leaves a vector invariant, the vector must be parallel (or, obviously, antiparallel) to the axis of rotation. Hence any given triplet VEV will conserve the subgroup formed by the rotations around the axis defined by the VEV. Thus the VEV of any triplet will break $SU(2)$ down to Spin(2), the double covering of $SO(2)$, which is in fact isomorphic to $SO(2)$ and $U(1)$.

Note that there is an $SO(2)$ for each possible axis, i.e.\ infinitely many $SO(2)$'s that are all mutually disjoint (up to the identity element). If we introduce two triplets, their VEVs will either be linearly dependent or not. If they are linearly dependent they break to the same subgroup. If they are linearly independent they break to disjoint subgroups, hence fully breaking $SO(3)$. As the triplet is an unfaithful representation of $SU(2)$, we will always conserve a subgroup $Z_2$ under which all components of the triplet transform trivially, while both components of the doublet transform non-trivially.

We conclude from this that if we use three-dimensional representations to break $SU(2)$, we either leave invariant a $U(1) \cong SO(2)$ symmetry or a $Z_2$. In particular no non-abelian subgroups can be conserved. We therefore do not need to consider combining a triplet VEV with a VEV of a different representation.

\mathversion{bold}
\subsection{Breaking $SU(3)$ with a triplet}
\mathversion{normal}

The argumentation for $SU(3)$ is in fact very similar to the one for $SU(2)$ broken by a triplet. As the intuitive geometric derivation used above is not so readily applied in the complex three-dimensional Euclidean space, we give a more elaborate proof using linear algebra. This derivation (with the obvious modifications) can also be applied to $SU(2)$.

In the three-dimensional representation of $SU(3)$ the group elements are mapped onto the 3$\times$3 unitary matrices with unit determinant. Therefore each element will have three eigenvalues $\lambda_1$, $\lambda_2$ and $\lambda_3$. If one of these eigenvalues, say $\lambda_1$, is 1 (i.e.\ if the element is able to be part of some conserved subgroup), then the other two eigenvalues have to fulfill $\lambda_2 \lambda_3 = 1$, since the matrix has a unit determinant. This means that if $\lambda_2$ is also equal to 1, then $\lambda_3=1$ as well. That is, the only element with more than one eigenvalue equal to 1 is the identity element, the only element with three 1 eigenvalues.\\
This means that each element which is not the identity will have at most one eigenvector corresponding to an eigenvalue of 1. For a simple example, the matrix
\begin{equation}
\left( \begin{array}{ccc}
e^{i \phi} & 0 & 0 \\
0 & e^{-i \phi} & 0 \\
0 & 0 & 1
       \end{array}
\right)
\end{equation}
\noindent will have the eigenvector
\begin{equation}
\left( \begin{array}{c}
0 \\
0 \\
1
       \end{array}
\right)
\end{equation}
\noindent corresponding to an eigenvalue of 1. As no direction in three-dimensional complex space is favored, there will exist for each complex 3-vector non-trivial group elements having this vector as an eigenvector with eigenvalue 1. These elements form the subgroup conserved by a VEV proportional to that eigenvector. As each non-trivial element has at most one such eigenvector, these subgroups will all be disjoint.

What is the subgroup conserved by such a VEV? We can already guess that it will be $SU(2)$, but this can be motivated by considering the group of elements that leave invariant a vector $\vec{v}$. We then make a unitary similarity transformation
\begin{equation}
U \rightarrow U'= \left( \begin{array}{ccc} \vec{x} & \vec{y} & \vec{v} \end{array} \right)^{\dagger} U \left( \begin{array}{ccc} \vec{x} & \vec{y} & \vec{v} \end{array} \right),
\end{equation}

\noindent where $U$ is an element of the group and $\vec{x}$ and $\vec{y}$ are arbitrary mutually orthogonal vectors that are also orthogonal to $\vec{v}$. We obtain
\begin{equation}
U' = \left( \begin{array}{cc} U'_{2 \times 2} & 0 \\ 0 & 1 \end{array} \right).
\end{equation}

\noindent As $U'$ is unitary by itself and also has unit determinant, we see that the three-dimensional representation reduces to the two-dimensional plus the one-dimensional representation of $SU(2)$. Since all the $SU(2)$ subgroups are disjoint, introducing two or more triplet scalars either breaks to an $SU(2)$ (in case their VEVs are linearly dependent) or breaks the entire $SU(3)$ group (if they are not).

What about anti-triplets? The arguments are the same as for the triplets, if we consider them separately, as the two representations can only be distinguished if they show up together. But even if we introduce scalars transforming as triplets and scalars transforming as anti-triplets, we do not find any new subgroups: The reasoning is the same as above, each scalar VEV breaks to a specific $SU(2)$ and they are all disjoint. The only thing we observe is that if we introduce a scalar triplet and a scalar anti-triplet they break to the same $SU(2)$ if the VEV of the triplet is proportional to the complex conjugated VEV of the anti-triplet. If this is not the case, they break to disjoint $SU(2)$'s, i.e.\ they fully break $SU(3)$.

We conclude that an arbitrary collection of scalar triplets and anti-triplets either conserves an $SU(2)$ subgroup of our original $SU(3)$ symmetry or fully breaks that symmetry.

\mathversion{bold}
\subsection{Breaking $SU(3)$ with the adjoint representation}
\mathversion{normal}

For discussing the breaking of a continuous group with matrix representations, we begin with the eight-dimensional adjoint representations of $SU(3)$, as it is probably the best known. We can write the VEV of a scalar transforming under the adjoint representation of $SU(3)$ as a Hermitian $3\times 3$ traceless matrix $V$. It then transforms under $SU(3)$ in the following way:
\begin{equation}
V \rightarrow V' = U V U^{\dagger},
\end{equation}

\noindent where $U$ is a special, unitary matrix. As $V$ is traceless, we need to consider two distinct cases: Either $V$ has three distinct eigenvalues, or it has two degenerate eigenvalues $\lambda$, the third eigenvalue being $-2 \lambda$. The only possible VEV with three degenerate eigenvalues is the zero matrix, i.e.\ a vanishing VEV, which naturally does not break $SU(3)$.

We first consider the case of a $V$ with three distinct eigenvalues. We are looking for the subgroup of $SU(3)$ formed by those elements $U$ which leave $V$ invariant, i.e.\ for which $V=V'$. This set is just the set of all matrices $U$ that commute with $V$. What does it mean if $U$ commutes with $V$? Let $\vec{v_i}$ be the eigenvector $V$ associated with the eigenvalue $\lambda_i$, which we have assumed to be nondegenerate. Then
\begin{equation}
V (U \vec{v_i})= U (V \vec{v_i}) = \lambda_i (U \vec{v_i}).
\end{equation}

\noindent Hence $U \vec{v_i}$ is also an eigenvector of $V$ with eigenvalue $\lambda_i$. As this eigenvalue is non-degenerate $U \vec{v_i}$ must linearly depend on $\vec{v_i}$. Therefore $\vec{v_i}$ is also an eigenvector of $U$. This holds for all three eigenvectors of $V$. We can thereby specify the subgroup conserved by this VEV: It is the set of all $U$ having the same set of eigenvectors as $V$. The most general form for an element of this group is then
\begin{equation}
\label{eq:Udecomp}
U= \left( \begin{array}{ccc} \vec{v_1} & \vec{v_2} & \vec{v_3} \end{array} \right)
\left( \begin{array}{ccc} e^{i \alpha} & 0 & 0 \\
0 & e^{i \beta} & 0 \\
0 & 0 & e^{-i (\alpha + \beta)}
       \end{array} \right)
 \left( \begin{array}{ccc} \vec{v_1} & \vec{v_2} & \vec{v_3} \end{array} \right)^{\dagger}.
\end{equation}

\noindent This representation is clearly unitarily equivalent to a diagonal representation, i.e.\ it reduces to three representations of $U(1)$. As $\alpha$ and $\beta$ are however independent, there are actually two distinct $U(1)$ groups. Therefore an adjoint VEV with three distinct eigenvalues breaks $SU(3)$ down to $U(1) \times U(1)$. Note that such a VEV can never conserve a non-abelian subgroup of $SU(3)$ and we do not need to consider it any further.

We now proceed to VEVs $V$ having two degenerate eigenvalues. The eigenvectors of $V$ are now no longer uniquely defined. If $\vec{v_1}$ and $\vec{v_2}$ are two orthonormal eigenvectors of $V$ corresponding to the same eigenvalue, we can find an arbitrary orthonormal basis of the corresponding eigenspace as $a \vec{v_1} + b \vec{v_2}$ and $- \overline{b} \vec{v_1} + \overline{a} \vec{v_2}$, with $a$ and $b$ two complex numbers obeying $\vert a \vert^2 + \vert b \vert^2 = 1$. Defining $X \equiv \left( \begin{array}{ccc} (a \vec{v_1} + b \vec{v_2}) & (- \overline{b} \vec{v_1} + \overline{a} \vec{v_2}) & \vec{v_3} \end{array} \right)$, we can therefore write any matrix in SU(3) that commutes with $V$ in the following form:
\begin{equation}
\label{eq:octet2degev}
X
\left( \begin{array}{ccc} e^{i \alpha} & 0 & 0 \\
0 & e^{i \beta} & 0 \\
0 & 0 & e^{-i (\alpha + \beta) }
       \end{array} \right)
X^{\dagger}
\end{equation}
\noindent To reduce this representation, we do a unitary equivalence transformation by multiplying on the right by
\begin{equation}
\left( \begin{array}{ccc} \vec{v_1} & \vec{v_2} & \vec{v_3} \end{array} \right)
\end{equation}
\noindent and on the left with the Hermitian conjugate. The resulting matrix is
\begin{equation}
\left( \begin{array}{ccc} 
\vert a \vert^2 e^{i \alpha} + \vert b \vert^2 e^{i \beta} & a \overline{b} (e^{i \alpha} - e^{i \beta}) & 0 \\
\overline{a} b (e^{i \alpha} - e^{i \beta}) & \vert a \vert^2 e^{i \beta} + \vert b \vert^2 e^{i \alpha} & 0 \\
0 & 0 & e^{-i (\alpha + \beta)} 
       \end{array} \right). \label{abcd}
\end{equation}

\noindent We now show that this is a representation of $SU(2) \times U(1)$. To do this we factorize the matrix (\ref{abcd}):
\begin{equation}
\left( \begin{array}{ccc} e^{i \frac{(\alpha+\beta)}{2}} & 0 & 0 \\
0 & e^{i \frac{(\alpha+\beta)}{2}} & 0 \\
0 & 0 & e^{-i (\alpha + \beta)}
       \end{array} \right)
\left( \begin{array}{ccc} 
x & y & 0 \\
-y^{*} & x^{*} & 0 \\
0 & 0 & 1 \end{array} \right),
\end{equation}
where
\begin{eqnarray}
x & = & \vert a \vert^2 e^{i \frac{(\alpha-\beta)}{2}} + \vert b \vert^2 e^{-i \frac{\alpha - \beta}{2}}, \label{xxx} \\
y & = & 2 i a b^{\ast} \sin{\left( \frac{\alpha - \beta}{2} \right) }. \label{yyy}
\end{eqnarray}
\noindent These two matrices commute. The first matrix is the representation of U(1), with the first two generations transforming in the same way, and the third with double and opposite charge. If we observe that $\vert x \vert^2 + \vert y \vert^2 =1$, we see that the second matrix furnishes a representation of SU(2), under which the first two generations form a doublet and the third generation is a singlet. We also note that we have the correct number of free parameters: The absolute value of $a$ (or $b$), the phase of $a b^{\ast}$ and the phase difference $(\alpha - \beta)$.

We consider the case of two adjoint VEVs, where both VEVs have degenerate eigenvalues. First of all, their non-degenerate eigenvalues could correspond to the same eigenvector. In this case, they will naturally break to the same subgroup. Then we could have the case, where the non-degenerate eigenvalue of the second VEV corresponds to an eigenvector lying in the eigenspace of the degenerate eigenvalue of the first VEV. This, in a way, singles out a basis of that eigenspace and thereby coincides with the VEV of an octet with three distinct eigenvalues, i.e.\ conserves a subgroup $U(1) \times U(1)$. Therefore, if there is no relation between the eigenvectors of the two VEVs, we only conserve the subgroup $Z_3$, corresponding to the three third roots of unity, which can never be broken by adjoint scalars. 

Finally, combining a degenerate adjoint VEV with a triplet VEV, we find three possibilities: First, the triplet VEV can coincide with the non-degenerate eigenvector. In this case $e^{-i (\alpha + \beta)}$ must be equal to 1 and we break down to the same $SU(2)$ conserved by the triplet VEV alone. If the triplet VEV lies in the degenerate eigenspace, we break the $SU(2)$ conserved by the octet VEV and are left with only a residual $U(1)$. If the triplet VEV is not an eigenvector of the adjoint VEV we again break the entire group.

Thus, the only new non-abelian subgroup of $SU(3)$ we can conserve with the VEV of a scalar transforming under the adjoint representation is the subgroup $SU(2) \times U(1)$ if the VEV has two degenerate eigenvalues.

\mathversion{bold}
\subsection{Breaking $SU(2)$ with the five-dimensional representation}
\mathversion{normal}

The calculations here are very similar to those of the last section, so we will be brief. The VEV $V$ of a scalar transforming under the five-dimensional representation can be written as a $3 \times 3$ traceless, real symmetric matrix. It transforms under $SU(2)$ as
\begin{equation}
V \rightarrow V' = O V O^T,
\end{equation}

\noindent with $O$ a special orthogonal matrix. Again the question of invariance can be reduced to a question of commutation and hence coincident eigenspaces. For a VEV with nondegenerate eigenvalues the general form of elements in the conserved subgroup is
\begin{equation}
O= \left( \begin{array}{ccc} \vec{v_1} & \vec{v_2} & \vec{v_3} \end{array} \right)
\left( \begin{array}{ccc} (-1)^n & 0 & 0 \\
0 & (-1)^m & 0 \\
0 & 0 & (-1)^{n+m}
       \end{array} \right)
 \left( \begin{array}{ccc} \vec{v_1} & \vec{v_2} & \vec{v_3} \end{array} \right)^T,
\end{equation}

\noindent with $n$, $m$ integers (as $V$ is symmetric it can be diagonalized by a real orthogonal matrix, hence $O$ will have only real eigenvectors and therefore only real eigenvalues). After a similarity transformation this is a representation of $Z_2 \times Z_2 \cong D_2$. However, since we actually break $SU(2)$ with an unfaithful representation, we actually conserve the double-valued group $D_2'$. The $SU(2)$ doublet will transform as a doublet in this group as well, while the triplet, as can be seen from the matrix above, decomposes into the three non-trivial one-dimensional representations. $D_2'$ has no non-abelian subgroups, so we do not need to consider a combination of this VEV with others.

Next we consider VEVs $V$ with two degenerate eigenvalues. The elements of the conserved subgroup must still have $\vec{v_3}$ as an eigenvector with a real eigenvalue. There are, however, two possibilities to do this. One is to assign the eigenvalue 1 to $\vec{v_3}$. These are all elements having $\vec{v_3}$ as axis of rotation. They form $SO(2)$ subgroup. We also have those elements, where the eigenvalue $-1$ is assigned to $\vec{v_3}$. These are of the form
\begin{equation}
Y \left( \begin{array}{ccc} (-1)^n & 0 & 0 \\
0 & (-1)^{n+1} & 0 \\
0 & 0 & -1
       \end{array} \right)
Y^T, \label{abcdef}
\end{equation}
\noindent using $Y \equiv \left( \begin{array}{ccc} (c \vec{v_1} + s \vec{v_2}) & (- s \vec{v_1} + c \vec{v_2}) & \vec{v_3} \end{array} \right)$, where $s$ and $c$ are the sine and cosine, respectively, of some undefined angle. By multiplying (\ref{abcdef}) on the right by
\begin{equation}
\left( \begin{array}{ccc} \vec{v_1} & \vec{v_2} & \vec{v_3} \end{array} \right),
\end{equation}

\noindent and on the left by its transpose, we perform a unitary transformation and end up with
\begin{equation}
\left( \begin{array}{ccc}
(-1)^n (c^2 - s^2)  & 2 c s (-1)^n  & 0 \\
2 c s (-1)^n  & (-1)^n (s^2 - c^2) & 0 \\
0 & 0 & -1
       \end{array} \right).
\end{equation}

\noindent Since the above matrix must still have a unit determinant, we know that the upper left $2 \times 2$ matrix must have determinant $-1$ and must also be orthogonal. Combining the two sets of elements, we find that our representation is reducible, reducing to the defining representation of $O(2)$ and the one-dimensional representation, where each element is mapped onto its determinant. As our original group was $SU(2)$, we are actually breaking to the double covering of $O(2)$, which is the group Pin(2). Combining several such VEVs, they can coincide in the non-degenerate eigenvector, in which case Pin(2) is conserved, the non-degenerate eigenvector of one can lie in the degenerate eigenspace of the other, in which case the conserved subgroup is $D_2'$, or their eigenbases could be unrelated, in which case only $Z_2$ is conserved.

There are thus only two non-abelian groups which can be the residual subgroup of $SU(2)$ after breaking with the VEV of a five-dimensional representation: The group $D_2'$ for non-degenerate eigenvalues and the group Pin(2) for degenerate eigenvalues. Some of these results can also be found in \cite{subgroups}.
\mathversion{bold}
\subsection{Breaking $SU(3)$ with the six-dimensional representation}
\mathversion{normal}

Writing the VEV of the six-dimensional representation as a complex, symmetric $3 \times 3$ matrix $V$, it transforms under $SU(3)$ in the following way:
\begin{equation}
V \rightarrow V' = U V U^T.
\end{equation}

\noindent Demanding invariance can then be rewritten as the condition
\begin{equation}
\label{eq:complexcomm}
UV=VU^{\ast}.
\end{equation}

\noindent We now note that $V$ need not necessarily be diagonalizable. However, since $V$ is complex and symmetric can be written in the form
\begin{equation}
W^T V W = V_{diag},
\end{equation}

\noindent with $W$ unitary \cite{Grimus:1988qr}. We can write $W \equiv \left( \vec{w_1}, \vec{w_2}, \vec{w_3} \right)$. The $\vec{w_i}$ are then singular vectors of $V$ obeying the relation
\begin{equation}
\label{eq:singvec}
V \vec{w_i} = \sigma_i \vec{w_i}^{\ast},
\end{equation}

\noindent with $\sigma_i$ the diagonal elements of $V_{diag}$, i.e.\ the singular values of $V$. The condition of equation~\eqref{eq:complexcomm} then leads to
\begin{equation}
V (U^{\ast} \vec{w_i}) = U V \vec{w_i} = \sigma_i U \vec{w_i}^{\ast} = \sigma_i (U^{\ast} \vec{w_i})^{\ast}.
\end{equation}

\noindent If $V$ has three distinct singular values, this means that all $\vec{w_i}$ need to be eigenvectors of $U^{\ast}$. Also, the corresponding eigenvalue of $U^{\ast}$ needs to be real. Therefore the discussion is the same as for the quintuplet of $SO(3)$: The conserved subgroup is $D_2$. If $V$ has two degenerate eigenvalues, then $U^{\ast}$ should act on the corresponding singular space with only real coefficients, that is it should be block-diagonalizable to give an orthogonal $2 \times 2$ submatrix. The conserved subgroup is then $O(2)$. As $V$ need not to be traceless, we encounter the additional case of three degenerate singular values. Here $U^{\ast}$ needs to act on all singular vectors with real coefficients, so the conserved subgroup in this case is $SO(3)$. Of these subgroups only the last two are non-abelian and need to be considered in combination with other VEVs.

We demanded above that the eigenvalues of $U$ need to be real. This condition stems from equation~\eqref{eq:singvec}: If $\vec{w_i}$ obeys that relation, then $\alpha \vec{w_i}$ only obeys the same relation if $\alpha$ is real or alternatively $\sigma_i$ must be zero. Thereby VEVs with zero eigenvalues are algebraically special: The group elements preserving this VEV can have complex eigenvalues corresponding to the singular vectors of $V$ with singular value 0. A special unitary matrix cannot have only one non-real eigenvalue. Hence the case of interest is a VEV with two zero eigenvalues. The group elements preserving this VEV are of the same form as those of equation~\eqref{eq:octet2degev}, with the additional condition that $e^{-i (\alpha + \beta) }$, the eigenvalue corresponding to the non-zero singular value, must be real. We can therefore substitute the parameter $\alpha + \beta$ by the integer parameter $m$ defined by $\alpha+\beta= m \pi$ and using definitions (\ref{xxx}, \ref{yyy}) write the group elements conserving $V$ in the form
\begin{equation}
\label{eq:SU2xZ4}
\left( \begin{array}{ccc} i^m & 0 & 0 \\
0 & i^m & 0 \\
0 & 0 & (-1)^m
       \end{array} \right)
\left( \begin{array}{ccc} 
x & y & 0 \\
-y^{*} &
x^{*} & 0 \\
0 & 0 & 1
       \end{array} \right).
\end{equation} 

\noindent The conserved subgroup is therefore $SU(2) \times Z_4$, where the first two generations form a doublet of $SU(2)$ and a faithful representation of $Z_4$, while the third generation is a singlet of $SU(2)$ and an unfaithful, non-trivial representation of $Z_4$.

What if we combine two six-dimensional VEVs? If they coincide in all three singular vectors, the subgroup is determined by the VEV with less degenerate eigenvalues. If they have only one singular vector in common, we break to the subgroup of elements having two degenerate real eigenvalues, that is $Z_2$. If they have no singular vectors in common, we break $SU(3)$ fully. Zero eigenvalues are only relevant if the VEVs coincide in all three singular vectors anyway and the zero eigenvalues correspond to the same eigenspace. In this case the full subgroup $SU(2) \times Z_4$ is conserved.

We thus have three non-abelian groups that can be conserved by a sextet VEV, $O(2)$ for two degenerate singular values, $SO(3)$ for three degenerate singular values, and $SU(2) \times Z_4$ for two zero eigenvalues. We now need to consider combinations of these three cases with the other VEVs we have discussed so far, triplets and octets.

What if both a $\Rep{6}$ and a triplet acquire a VEV? If the triplet VEV is not a singular vector of the $\Rep{6}$, then $SU(3)$ is fully broken. What if it is a singular vector? If $V$ has two degenerate singular values, the triplet can correspond to the non-degenerate singular value. In this case, the determinant of the $2 \times 2$ submatrix is fixed to be one, and the conserved subgroup is $SO(2)$ or $U(1)$. If the triplet VEV corresponds to a degenerate singular value, the degeneracy becomes irrelevant and the subgroup is $Z_2$. If $V$ has three degenerate singular values, the triplet, which is in the defining representation of $SO(3)$, breaks that subgroup in the usual way down to $U(1)$ or fully breaks it, if the real and imaginary parts of the triplet VEV are not parallel. If we combine a $V$ with two zero eigenvalues with a triplet VEV, we again have two possibilities: The triplet VEV can correspond to the non-zero singular value. In this case $m$ is fixed to be 0 or 2, and we break down to $SU(2)$ (the former $Z_4$ element can just be multiplied into the $SU(2)$ element, without changing the determinant). If the triplet VEV is an eigenvector of $V$ corresponding to a zero eigenvalue, we need to take a closer look. The $SU(2)$ element in equation~\eqref{eq:SU2xZ4} has eigenvalues $e^{\pm i \frac{(\alpha-\beta)}{2}}$. So, without loss of generality, we must now demand $(i)^m e^{i \frac{(\alpha-\beta)}{2}} = 1$. The resulting element then has in addition two eigenvalues of $-1$, corresponding to fixed vectors. The conserved subgroup is then $Z_2$.

We proceed by combining a VEV of a $\Rep{6}$ with an adjoint VEV. The adjoint VEV must have two degenerate eigenvalues, as only then do we have the possibility of conserving a non-abelian subgroup. If there does not exist a basis of singular vectors for the sextet VEV, that is also a basis of eigenvectors for the octet VEV, $SU(3)$ will be fully broken. In particular, the eigenvector of the octet VEV corresponding to the non-degenerate eigenvalue, $\vec{v_3}$, must always be a singular vector of the sextet VEV $V$. The discussion is then similar to that for the triplet, with the triplet VEV replaced by $\vec{v_3}$. If $V$ has two degenerate singular values, $\vec{v_3}$ can correspond to the non-degenerate singular value. In this case nothing changes and the conserved subgroup is $O(2)$. If $\vec{v_3}$ corresponds to a degenerate singular value, the degeneracy becomes irrelevant and the subgroup is $D_2$. If $V$ has three degenerate singular values, one of the degeneracies becomes irrelevant and we break down to $O(2)$. Finally, considering the case of a sextet VEV $V$ with two zero eigenvalues, we again have two possibilities: $\vec{v_3}$ can correspond to the non-zero singular value. In this case nothing changes, and  $SU(2) \times Z_4$ is still the conserved subgroup. If $\vec{v_3}$ is an eigenvector of $V$ corresponding to a zero eigenvalue, a specific basis is singled out for the elements of the conserved subgroup. It is thus only determined by the possible eigenvalues, and cannot be non-abelian. In this case it will be $U(1) \times Z_2$. Thus, no new non-abelian subgroups can be attained by combining the VEVs of these different $SU(3)$ representations.

\mathversion{bold}
\subsection{Breaking $SU(2)$ with the four-dimensional representation}
\mathversion{normal}

We finally deal with the most complicated of the matrix representations, the $\Rep{4}$ of $SU(2)$. As it arises from the product of a vector and a spinor, it can be written as a $3 \times 2$ complex matrix, with one spinor index and one vector index. There must be further constraints, as such a matrix has 6 complex degrees of freedom. To find them, we take a look at the Clebsch Gordan coefficients.

Writing the $\Rep{4}$ as a matrix, it acts on a spinor and transforms it into a vector. As the Clebsch Gordan coefficients are normally given in spherical coordinates we start with these, later switching back to Cartesian coordinates, where the scalar product of two vectors is simply matrix multiplication. In spherical coordinates, we can give the four degrees of freedom of the $\Rep{4}$ as $\phi_1$ (m=$\frac{3}{2}$), $\phi_2$ (m=$\frac{1}{2}$), $\phi_3$ (m=$-\frac{1}{2}$) and $\phi_4$ (m=$-\frac{3}{2}$). Correspondingly we write the two components of the spinor we want to transform into a vector as $\psi_1$ (m=$\frac{1}{2}$) and $\psi_2$ (m=$-\frac{1}{2}$). Using the Clebsch Gordan coefficients for $SU(2)$ \cite{Amsler:2008zzb} we find that they combine in the following way to form a vector:
\begin{eqnarray}
\frac{1}{2} (\sqrt{3} \phi_1 \psi_2 - \phi_2 \psi_1) & (m = & 1),\\
\frac{1}{\sqrt{2}} (\phi_2 \psi_2-\phi_3 \psi_1) & (m= & 0), \\
\frac{1}{2} (\phi_3 \psi_2-\sqrt{3} \phi_4 \psi_1) & (m= & -1). 
\end{eqnarray}

\noindent Switching to Cartesian coordinates, this corresponds to a vector
\begin{equation}
\left( \begin{array}{c} 
\frac{1}{2 \sqrt{2}} [(\phi_2 - \sqrt{3} \phi_4) \psi_1 + (\phi_3-\sqrt{3} \phi_1) \psi_2] \\
\frac{i}{2 \sqrt{2}} [-(\phi_2 + \sqrt{3} \phi_4) \psi_1 + (\phi_3+\sqrt{3} \phi_1) \psi_2)] \\
\frac{1}{\sqrt{2}} (\phi_2 \psi_2-\phi_3 \psi_1)
       \end{array} \right).
\end{equation}

\noindent This vector arises from multiplying a spinor by the following matrix:
\begin{equation}
\frac{1}{\sqrt{2}} \left( \begin{array}{cc}
\frac{1}{2} (\phi_2 - \sqrt{3} \phi_4) & \frac{1}{2} (\phi_3-\sqrt{3} \phi_1) \\
-\frac{i}{2} (\phi_2 + \sqrt{3} \phi_4)& \frac{i}{2} (\phi_3+\sqrt{3} \phi_1) \\
-\phi_3 & \phi_2
       \end{array} \right),
\end{equation}

\noindent or, in another simpler parameterization
\begin{equation}
V=\left( \begin{array}{cc} a & b \\ c & d \\ -b+id & a+ic \end{array} \right),
\end{equation}

\noindent where $a$, $b$, $c$ and $d$ are complex. This is then the most general form for the VEV of a $\Rep{4}$. It transforms in the following way:
\begin{equation}
\label{eq:4cond}
V \rightarrow V' = O V U^{\dagger},
\end{equation}

\noindent as it has one vector and one spinor index. $O$ and $U$ are of course not independent but describe a rotation of the same magnitude around the same axis. It can be checked by explicit calculation that $V'$ can be parameterized in the same way as $V$ for an arbitrary rotation.

Again we reformulate the condition of invariance as a condition on the eigensystems. We first observe that we can deduce from equation~\eqref{eq:4cond} the following two equations:
\begin{eqnarray}
V V^{\dagger} &=& O V V^{\dagger} O^T \\
V^{\dagger} V &=& U V^{\dagger} V U^{\dagger},
\end{eqnarray}

\noindent from which we immediately deduce that the eigenvectors of $VV^{\dagger}$ (i.e.\ the left singular vectors of $V$, denoted by $\vec{u_i}$) must also be eigenvectors of $O$ (with the usual ambiguities for degenerate singular and eigenvalues) and the right singular vectors of $V$, denoted by $\vec{w_i}$,  must be eigenvectors of $U$. Using this knowledge, we go back to equation~\eqref{eq:4cond}. We find that
\begin{equation}
V U \vec{w_i} = V \mu_i \vec{w_i} = \sigma_i \mu_i \vec{u_i},
\end{equation}

\noindent for $i=1,2$, $\mu_i$ the eigenvalues of $U$ and $\sigma_i$ the singular values of $V$. We also have
\begin{equation}
V U \vec{w_i} = O V \vec{w_i} = O \sigma_i \vec{u _i} = \lambda_i \sigma_i \vec{u_i},
\end{equation}

\noindent with $\lambda_i$ the eigenvalues of $O$. From the last two equations we can deduce that $\lambda_i = \mu_i$. As $O$ and $U$ are rotations by the same angle $\theta$, their eigenvalues are $e^{\pm i \theta}$ and 1 for $O$ and $e^{\pm i \frac{\theta}{2}}$ for $U$. How can they be made to coincide? Apart from the trivial case of both being the unit matrix, we are only left with the possibility of identifying the exponential eigenvalues, which is only possible for $\theta= \pm \frac{4 \pi}{3}$. The final left singular vector $\vec{u_3}$ is then the eigenvector of $O$ corresponding to the eigenvalue 1, i.e.\ it defines the axis of rotation. If it is real, we then break to all rotations around that axis, with the angles given above. This is a $Z_3$ subgroup of $SU(2)$. If the axis is complex (and real and imaginary parts are not linearly dependent), no such elements exist and $SU(2)$ is fully broken.

Is the subgroup enlarged if $V$ has degenerate singular values? We first take the case $\sigma = \sigma_1 = \sigma_2 \neq 0$. $\vec{u_3}$ is still an eigenvector of $O$, the $\vec{u_i}$ and $\vec{w_i}$ however need not be eigenvectors of $O$ and $U$, respectively. Rather we have
\begin{eqnarray}
V U \vec{w_i} = & V (\alpha_i \vec{w_1} + \beta_i \vec{w_2}) &= \sigma (\alpha_i \vec{u_1} + \beta_i \vec{u_2}) \\
V U \vec{w_i} = & O V \vec{w_i} = \sigma O \vec{u_i} &= \sigma (\alpha'_i \vec{u_1} + \beta'_i \vec{u_2}),
\end{eqnarray}

\noindent from which we can immediately infer that $\alpha_i = \alpha'_i$ and $\beta_i = \beta'_i$. This means that again their eigenvalues need to coincide, and we again break to $Z_3$ or nothing. Finally, we need to consider the case, where one of the non-trivial singular values is zero, say $\sigma_2=0$. In this case $O$ and $U$ need only coincide in one eigenvalue, but this condition is already strong enough to constrain the elements in the same way, i.e.\ giving $Z_3$ as the conserved subgroup. We thus find that a VEV for the four-dimensional representation of $SU(2)$ can never lead to the conservation of a non-abelian subgroup, and we thus do not need to combine it with any other VEVs.

\mathversion{bold}
\section{Conclusions}
\mathversion{normal}

\noindent The results of this paper can be summarized in one sentence: The only non-abelian discrete subgroup that can be conserved by VEVs of the small representations (dimension equal or less than five and eight, respectively) of $SU(2)$ and $SU(3)$ is the group $D_2'$.

To arrive of this result, we have assumed that the continuous flavor symmetry is very ``cleanly'' broken - what we mean by this, is that there are no low-mass remnants of the symmetry breaking floating around in the resulting low energy model, such as (Pseudo-) Nambu-Goldstone bosons (which can result either from the continuous symmetry itself, if it is not gauged, or from accidental symmetries of the symmetry-breaking potential) or low mass scalar degrees of freedom related with (approximately) flat directions of the symmetry-breaking potential, such as often appear in supersymmetric frameworks. This clean breaking is what is needed to reproduce models using a discrete non-abelian symmetry as starting point, as they only relate to a possible larger symmetry through embedding and consistency constraints.

It may, however, be possible to define intermediate non-abelian symmetries in a somewhat less strict sense. Here it would not be possible to describe an intermediate model with a discrete symmetry independently of the underlying continuous symmetry, but it can still be interesting to examine the role of the discrete symmetry in such models more closely. An example is the $A_4$ symmetry appearing in \cite{Stech:2008wd}.

What further implications do the results presented in this paper have for flavored model-building with non-abelian discrete flavor symmetries? Our results naturally still allow for more complicated schemes for breaking a continuous flavor symmetry to a discrete subgroup. The symmetry breaking fields would then not couple to the fermions at leading order and a dedicated study would be necessary to check whether a lower energy model with a discrete flavor symmetry could be reproduced.

Our results may be taken as an indication that one may have to think differently about intermediate discrete flavor symmetries. If one favors discrete flavor symmetries in their own right at some scale, one should look towards other possible origins, such as from extra dimensions, which would also protect such a global symmetry from quantum gravity effects that might otherwise eradicate it \cite{Krauss:1988zc}. The main lesson is that the embedding of a discrete flavor symmetry in a continuous one is not nearly as simple as is often tacitly assumed.

\section*{\label{sec:Ack}Acknowledgments} 

This work has been supported by the DFG-Sonderforschungsbereich Transregio 27 ``Neutrinos and beyond -- Weakly interacting particles in Physics, Astrophysics and Cosmology''.


\newpage

\begin{thebibliography}{99}

\bibitem{Kobayashi:2006wq}
  T.~Kobayashi, H.~P.~Nilles, F.~Ploger, S.~Raby and M.~Ratz,
  Nucl.\ Phys.\  B {\bf 768}, 135 (2007)
  [arXiv:hep-ph/0611020].

\bibitem{Altarelli:2006kg}
  G.~Altarelli, F.~Feruglio and Y.~Lin,
  Nucl.\ Phys.\  B {\bf 775}, 31 (2007)
  [arXiv:hep-ph/0610165].

\bibitem{Ourorbifold}
  A.~Adulpravitchai, A.~Blum and M.~Lindner,
  [arXiv:0906.0468 [hep-ph]].

\bibitem{subgroups}
  P.~H.~Frampton and A.~Rasin,
  Phys.\ Lett.\  B {\bf 478}, 424 (2000)
  [arXiv:hep-ph/9910522].

\bibitem{Hagedorn:2006ug}
  C.~Hagedorn, M.~Lindner and R.~N.~Mohapatra,
  JHEP {\bf 0606}, 042 (2006)
  [arXiv:hep-ph/0602244].

\bibitem{Tanimoto:1999pj}
  M.~Tanimoto, T.~Watari and T.~Yanagida,
  Phys.\ Lett.\  B {\bf 461}, 345 (1999)
  [arXiv:hep-ph/9904338].

\bibitem{Araki:2008ek}
  T.~Araki, T.~Kobayashi, J.~Kubo, S.~Ramos-Sanchez, M.~Ratz and P.~K.~S.~Vaudrevange,
  Nucl.\ Phys.\  B {\bf 805}, 124 (2008)
  [arXiv:0805.0207 [hep-th]].

\bibitem{Luhn:2008sa}
  C.~Luhn and P.~Ramond,
  JHEP {\bf 0807}, 085 (2008)
  [arXiv:0805.1736 [hep-ph]].

\bibitem{Li:1973mq}
  Ling-Fong Li,
  Phys.\ Rev.\  D {\bf 9}, 1723 (1974).

\bibitem{Etesi:1997jv}
  G.~Etesi,
  J.\ Math.\ Phys.\  {\bf 37}, 1596 (1996)
  [arXiv:hep-th/9706029].

\bibitem{Koca:1997td}
  M.~Koca, M.~Al-Barwani and R.~Koc,
  J.\ Phys.\ A  {\bf 30}, 2109 (1997).

\bibitem{Koca:2003jy}
  M.~Koca, R.~Koc and H.~Tutunculer,
  Int.\ J.\ Mod.\ Phys.\  A {\bf 18}, 4817 (2003)
  [arXiv:hep-ph/0410270].

\bibitem{Blum:2007jz}
  A.~Blum, C.~Hagedorn and M.~Lindner,
  Phys.\ Rev.\  D {\bf 77}, 076004 (2008)
  [arXiv:0709.3450 [hep-ph]].

\bibitem{Frigerio:2004jg}
  M.~Frigerio, S.~Kaneko, E.~Ma and M.~Tanimoto,
  Phys.\ Rev.\  D {\bf 71}, 011901 (2005)
  [arXiv:hep-ph/0409187].

\bibitem{Grimus:1988qr}
  W.~Grimus and G.~Ecker,
  J.\ Phys.\ A  {\bf 21}, 2825 (1988).

\bibitem{Amsler:2008zzb}
  C.~Amsler {\it et al.}  [Particle Data Group],
  Phys.\ Lett.\  B {\bf 667}, 1 (2008).

\bibitem{Stech:2008wd}
  B.~Stech and Z.~Tavartkiladze,
  Phys.\ Rev.\  D {\bf 77}, 076009 (2008)
  [arXiv:0802.0894 [hep-ph]].

\bibitem{Krauss:1988zc}
  L.~M.~Krauss and F.~Wilczek,
  Phys.\ Rev.\ Lett.\  {\bf 62}, 1221 (1989).



\end{thebibliography}
\end{document}